\documentclass[aps,prb,twocolumn,groupedaddress]{revtex4}
\usepackage[dvips]{graphics}
\usepackage{vmargin}
\usepackage[dvips]{color}

\setpapersize{USletter}
\setmarginsrb{1in}{1.5in}{1in}{0.5in}{0in}{0.2in}{0in}{0in}

\begin{document}

\title{Magnetic properties of Mn impurities on GaAs (110) surfaces}

\author{M. Fhokrul Islam}
\author{C. M. Canali}
\affiliation{School of Computer Science, Physics and Mathematics,
Linn\ae us University, 391 82 Kalmar,
Sweden}

\date{\today}

\begin{abstract}
We present a computational study of individual and pairs of
substitutional Mn impurities on the (110) surface of GaAs samples
based on density functional theory. We focus on the anisotropy
properties of these magnetic centers and their dependence on on-site
correlations, spin-orbit interaction and surface-induced
symmetry-breaking effects. For a Mn impurity on the surface, the
associated acceptor-hole wavefunction tends to be more localized
around the Mn than for an impurity in bulk GaAs. The magnetic
anisotropy energy for isolated Mn impurities is of the order of 1
meV, and can be related to the anisotropy of the orbital magnetic
moment of the Mn acceptor hole. Typically Mn pairs have their spin
magnetic moments parallel aligned, with an exchange energy that
strongly depends on the pair orientation on the surface. The spin
magnetic moment and exchange energies for these magnetic entities
are not significantly modified by the spin-orbit interaction, but
are more sensitive to on-site correlations. Correlations in general
reduce the magnetic anisotropy for most of the ferromagnetic Mn
pairs.
\end{abstract}


\maketitle

\section{Introduction}
\label{introduction}

Since the discovery of ferromagnetic order in Mn-doped GaAs with a
Curie temperature above of 100 K,\cite{ohno_1996} research in dilute
magnetic semiconductors (DMS) has developed into an important branch
of material science.  Early work, besides being aimed at
understanding the physics of (Ga, Mn)As was strongly focused on the
goal of achieving room-temperature ferromagnetism in this prototype
DMS. Although this goal seems now of difficult realization, (Ga,
Mn)As and DMSs in general still attract a lot of attention both for
fundamental science and applications (e.g. in
spintronics).\cite{ohno_nat_mat_2010, dietl_nat_mat_2010} From the
point of view of theory, intense effort based on both
phenomenological models\cite{dietl00, macdonald05, janko02, timm05,
hilbert05, jungwirth05,jungw_rmp06, zemen09, tang08, burch08} and ab-initio
calculations\cite{zunger1, zunger2, eriksson04, ebert09, sato10} has
lead to undoubted progress in understanding the origin and the
properties of ferromagnetism in (Ga, Mn)As. Yet, some fundamental
aspects related to the microscopic mechanism remain controversial
and are still strongly debated.\cite{masek_prl_2010,
ohya_nat_phys_2011,dietl_prb_2011}

On the experimental front, recent scanning tunneling microscopy
(STM) studies in (Ga, Mn) As have been able to visualize the spatial
structure of the acceptor wavefunctions at doping concentrations
near the metal-insulator transition and shown that it has a
multifractal character, possibly indicating carrier
correlations.\cite{richardella_science_2010} Similar STM experiments
have been carried in the last few years at very low doping
concentrations,\cite{yakunin04,yakunin05, yazdani06, koenraad_prb08,
yazdani09, celebi_koenraad_prl_2010, gupta_science_2010} probing
specifically the properties of isolated single and pairs of
substitutional Mn impurities in GaAs. Apart from helping understand
the very dilute limit of magnetic semiconductors, these studies are
interesting {\it per se},\cite{Koenraad2011} in that individual
magnetic impurities in semiconductors represent novel nano-magnetic
entities with unusual properties and promising
applications.\cite{Koenraad2011} The high-resolution STM
measurements provide detailed information on e.g. the character of
the wavefunctions of single magnetic dopants in semiconductors, and
the exchange interaction between two nearby isolated magnetic
impurities interacting via their associated itinerant holes.

Theoretical approaches based on tight-binding
models\cite{tangflatte_prl04, canali09} have so far provided a clear
picture of many of the STM experimental findings on single dopants,
including the anisotropic form of the Mn acceptor wavefunction and
its dependence on the Mn spin direction,\cite{tangflatte_prb05,
canali09} and the dependence of the exchange coupling between two Mn
impurities in GaAs on pair orientation and atomic
separation.\cite{tangflatte_prl04,yazdani06,tangflatte_spie_2009,
canali10} Tight-binding calculations are also able to provide a
description of how the properties of the Mn acceptor state depend on
the distance of the magnetic impurities from the surface
layer\cite{canali09}, which is in qualitative agreement with
experiment.\cite{garleff_prb_2010, gupta_science_2010}

First-principles calculations based on density functional theory
(DFT) play an important role in the theoretical study of
DMS.\cite{sato10} With the caveat of the well-known difficulties of
DFT of predicting correctly band gaps semiconductor and dealing with
localized strong correlations at the impurity sites, DFT-based
first-principle calculations have become a very powerful tool to
investigate the electronic structure and the magnetic properties of
DMS. For the specific case of the effective exchange coupling
between two isolated magnetic impurities in {\it bulk} GaAs, earlier
DFT calculations\cite{zunger1, zunger2} have provided useful
information on its microscopic origin and its strong anisotropic
character with respect to pair orientation. DFT techniques have also
been used to simulate STM images of the Mn acceptor
states,\cite{eriksson04, stroppa2007} both for interstitial and
substitutional impurities, yielding results that are qualitatively
consistent with experiment.\footnote{These calculations, although in
principle more accurate than the tight-binding calculations
mentioned above, have the drawback that the surface area that can be
simulated is rather small.}

A quantity that so far has not been thoroughly investigated by
first-principle methods is the magnetic anisotropy energy,
particularly for the case of magnetic impurities located close to
the symmetry-breaking surface which provides STM access. The
magnetic anisotropy is a very important quantity, particularly when
it comes to utilizing these nano-magnets in spintronics
applications. Indeed, the minima in the magnetic anisotropy
landscape determine the direction of the magnetization. In DMSs the
magnetic anisotropy barriers can be efficiently varied with an
external electric field, which can change the carrier concentration,
via the spin-orbit interaction (SOI).\cite{ohno_nat_mat_2010}
Therefore it is in principle possible to control and manipulate the
magnetization direction solely by means of an electric field.
Tight-binding calculations provide an estimate of the magnetic
anisotropy for Mn impurities on the GaAs surface.\cite{canali09}
Nevertheless, microscopic calculations based on first-principles
approach are certainly desirable.

In this paper we present first-principles calculations of the
magnetic properties for single and pairs of substitutional Mn
impurities on the (110) surface of GaAs, which is the most common
cleaved surface employed in cross sectional STM studies. We focus,
in particular on the anisotropic characteristics of important
magnetic quantities, such as total energy, spin and orbital moments
and exchange coupling, resulting from the interplay of SOI and
symmetry-breaking surfaces. Since on-site self-interaction
corrections on magnetic impurities are crucial to correctly describe
the Mn d electronic states in GaAs, we carry out our calculations in
the framework of the Generalized Gradient Approximation + Hubbard
$U$ (GGA + $U$) scheme, and we examine the effect of on-site
correlations on the anisotropic properties.

We find that the typical magnetic anisotropy energy for a single
substitutional Mn impurity on the (110) plane is of the order one 1
meV, with easy axis in the plane. These conclusions are only
slightly affected by the presence of the Hubbard $U$ term, which on
the other hand modifies the value of the total spin moment. The
calculation of the total orbital moment yields very small values,
ascribable to contributions coming predominately from the acceptor p
states and, to a lesser extent, from the Mn d states. For two nearby
Mn impurities, the magnetic anisotropy/atom is of the same order of
the single-impurity anisotropy. The effective exchange coupling
between the two Mn is typically ferromagnetic (i.e., the
energetically stable configuration has two moments aligned
parallel), and strongly anisotropic with respect to pair orientation
in the surface and atom separation. These anisotropies are mainly a
result of the GaAs crystal structure and symmetry-breaking surface.
The effect of the SOI on the exchange constant and spin moment is,
as expected, rather small.

On the other hand, the exchange coupling is affected by the Hubbard
$U$ term, particularly when the Mn atoms are at the shortest
possible separation. This effect persists at larger separations for
a Mn pair in the bulk, but it is significantly smaller for a Mn pair
on the surface, for which the Mn acceptor wavefunctions tend to be
more localized.

The paper is organized as follows. In Sec.~\ref{computation} we
discuss some technical aspects of the DFT calculations presented in
this paper. Sec.~\ref{results_section_1Mn} contains the results of
our numerical simulations for individual Mn impurities on a (110)
GaAs surface. The properties on a Mn pair, with a discussion of the
effective exchange interaction is presented in
Sec.~\ref{results_section_2Mn}. Finally
Sec.~\ref{conclusions_section} summarizes our work and discusses its
implications for the physics of magnetic impurities in
semiconductors and semiconductor spintronics.

\section{Computational details}
\label{computation}

The density functional theory (DFT) calculations in this work are
performed using the generalized gradient approximation (GGA) with
Perdew-Burke-Ernzerhof (PBE) exchange-correlation
functional.\cite{perdew96} For most part of the numerical
calculations the method of full potential Linearized Augmented Plane
Wave with local orbitals (LAPW+lo), as implemented in WIEN2k, is
used.\cite{wien2k} Because of the time-expensive nature of the plane
wave method, we have used the SIESTA\cite{siesta} ab-inito package
for relaxing the surfaces in our calculations, SIESTA employs pseudo
potentials and a numerical basis set. The relaxed coordinates are
then used as an input for the WIEN2k calculations.

\begin{figure}[h]
{\resizebox{2.5in}{3in}{\includegraphics{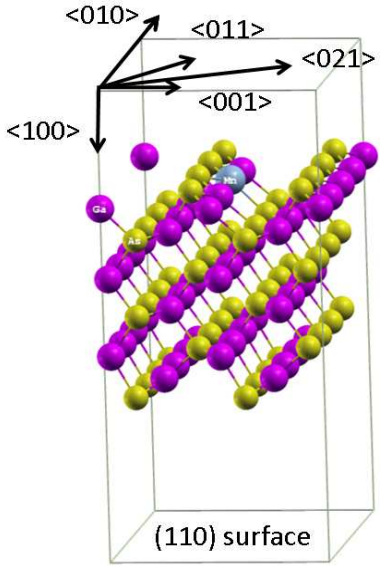}}} \caption{A
Relaxed GaAs (110) surface with a Mn impurity. The arrows indicate
different crystallographic axes of the surface supercell defined for
this work.} \label{surf_relx}
\end{figure}

The (110) surface of GaAs for our calculation is constructed by
cutting the bulk crystal along $<110>$ direction. The surface
supercell consists of six layers with 18 atoms at each layer ie
total of 96 atoms. A vacuum of 25 Bohr is added along the surface as
shown in Fig.~\ref{surf_relx}.

To study the effect of an isolated Mn impurity on the surface we
have replaced one Ga atom from the center of the top layer, as shown
in Fig.~\ref{surf_relx}. To calculate the anisotropy energy we have
calculated the effect of SOI for different magnetization directions,
namely $<100>, <001>, <011>$ and $<010>$ directions. The anisotropy
energy is the difference between the largest and the smallest
energies. The spin-orbit coupling in Wien2k is incorporated via a
second variational step using scalar relativistic eigenstates as
basis.\cite{so}

The strong correlation among the d electrons of Mn atoms is
accounted for by adding an orbital dependent potential $U$ for the d
shell electrons. The value of $U$ for Mn is usually chosen between 3
and 4 eV to match photoemission spectra and in our calculation we
have used $U= 4$ eV following K. Sato {\it et al.}~\cite{sato10}

To study the dependence of the exchange energy on the pair
orientation we replace two Ga atoms by two magnetic impurities. One
of the impurity atoms is kept fixed at the top layer of the
supercell (front left corner of Fig.~\ref{surf_relx}) and the
position of the second one is varied along different
crystallographic directions, namely [001], [010], [011] and [021].
In this paper we only consider the case in which the spin magnetic
moments of the two impurities are collinear. The exchange energy for
each pair orientation is calculated as the difference in energies of
the supercell when the spins of the impurity atoms are arranged
parallel and antiparallel. For the anisotropy calculation with two
impurities we have followed the same procedure as we have done for
the single impurity case.

Since the size of the surface supercell is large, we have used only
one k-point for these calculations and self-consistency is achieved
when total energies are converged to within 0.01 mRy or better and
charge is converged to 0.001e or better.

\section{Results}
\label{results_section}

\subsection{Single Mn on (110) GaAs surface}
\label{results_section_1Mn}

\subsubsection{Electronic structure}

\begin{figure}[htbp]
{\resizebox{3.2in}{2.2in}{\includegraphics{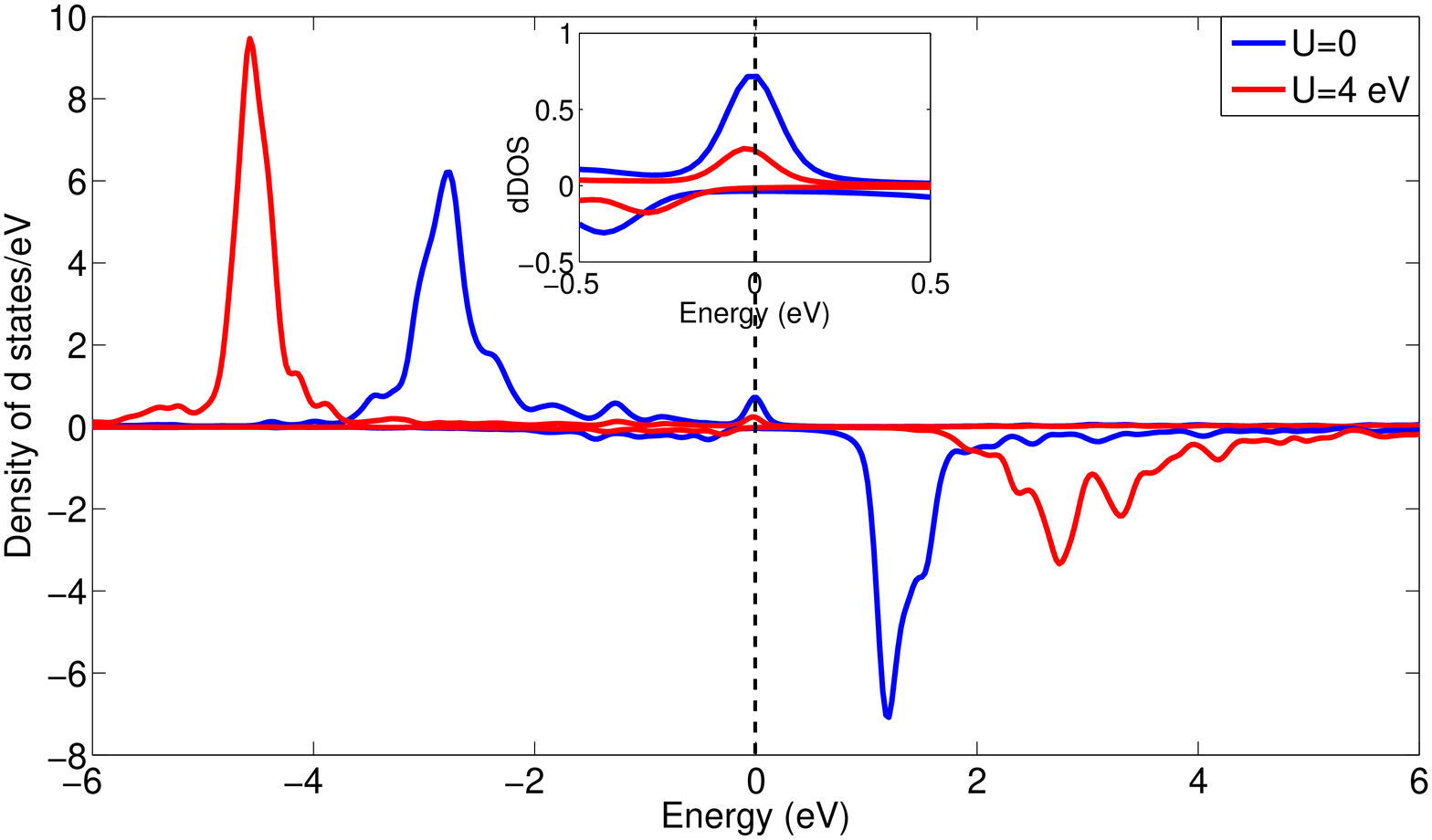}}}
\caption{(Color online) The density of Mn d states calculated with
$U$=0 (blue-dark line) and $U$=4 eV (red-gray line) respectively.
The inset shows the DOS near the Fermi level.} \label{d_dos}
\end{figure}

We start by analyzing the main features of the electronic structure
when an individual Mn impurity replaces a Ga atom on the (110)
surface of GaAs. A Mn atom introduces a magnetic moment in the host
material. This can be seen by plotting the partial Density of States
(DOS) for the Mn d-levels as shown in Fig.~\ref{d_dos}. Intra-atomic
exchange interactions, responsible for Hund's first rule, split the
majority (up-spin) from the minority (down-spin) d-states, so that
the former end up -- almost entirely -- below the Fermi level and
the latter above, thus giving rise to a localized spin moment in the
system. In Fig.~\ref{d_dos} we also show the effect of on-site
correlations on the d-orbital partial DOS, included via a Hubbard
$U$ term in the GGA calculations. Clearly correlations split
majority and minority states further apart, pushing the former
deeper below the Fermi level. As discussed below, correlations also
decrease the small peak in the partial DOS at the Fermi energy as
shown in the inset of Fig.~\ref{d_dos}.

\begin{figure}[htbp]
{\resizebox{3.4in}{2.7in}{\includegraphics{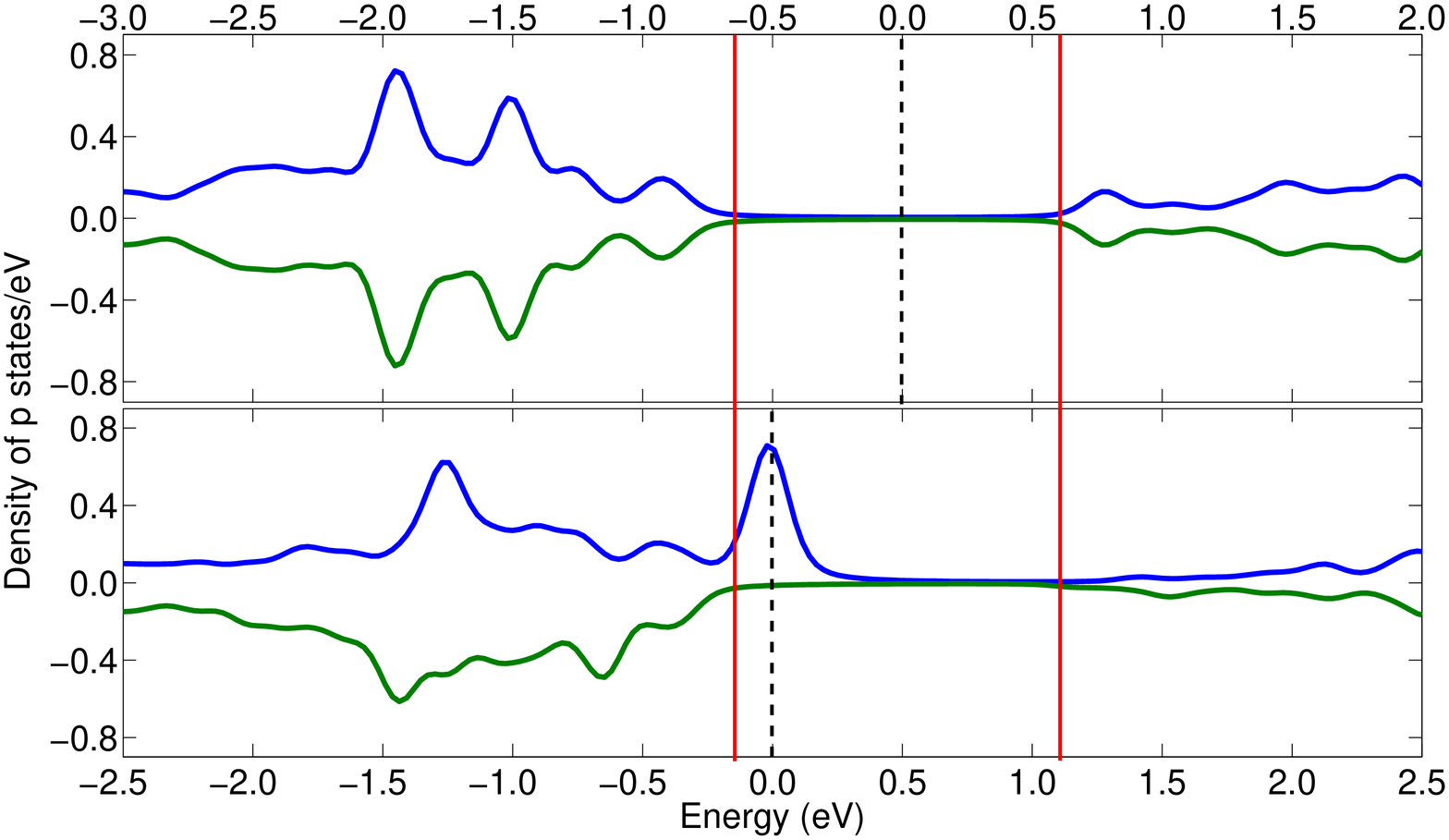}}}
\caption{(Color online)(Top) The density of As p states of pure GaAs
surface. (Bottom) The p partial DOS of the same As atom when a
nearby Ga atom is substituted by a Mn atom. The vertical dashed line
in both panels represents the Fermi energy. The two vertical red
lines delimit the energy-gap region for pure GaAs, which is clearly
identifiable in the top panel. To compare the two cases, we have
aligned the edge of the two conduction bands.} \label{p_dos}
\end{figure}

The second effect caused by a substitutional Mn impurity in GaAs is
the introduction of a p-like acceptor (hole) state in its
surroundings. This can be seen by plotting the p-orbital partial DOS
at the Mn nearest-neighbor As sites, as shown in Fig.~\ref{p_dos}.
As a comparison, in Fig.~\ref{p_dos}(a) the p-orbital partial DOS at
an As site is plotted for pure GaAs (i.e., when the Mn is replaced
by a Ga atom). There we can see the expected (full) valence and
(empty) conduction band, with a Fermi level in the middle of the
energy gap that separates them. When the Mn is present, the top of
the valence band spin polarizes as a result of the hybridization
between the As p orbitals with the Mn d-orbitals. In
Fig.~\ref{p_dos}(b) we can see that the spin-up component (i.e.,
parallel to the Mn spin moment) of the As p-orbital partial DOS has
a broad peak centered at the Fermi energy. The corresponding
spin-down component is instead pushed below the Fermi energy,
causing an effective exchange splitting between up and down states
of the order of $\sim$ 0.6 eV.

The spin-polarized states just above $E_F$ can be identified with
the acceptor hole states associated with the Mn impurity. These
states are clearly the result of a moderate hybridization between
spin polarized localized Mn d orbitals with p orbitals primarily
located on the neighbor As atoms. Note that the DOS of p character
is larger than the d contribution. This feature will have an impact
on the nature of the anisotropy energy discussed below. The inset of
Fig.~\ref{d_dos} shows that on-site correlations decrease the
majority $d$ partial DOS at the Fermi energy. As a result, the
hybridization of the hole state with the $d$ orbitals decreases, and
the acceptor becomes less localized on the Mn.

\begin{figure}[htbp]
{\resizebox{3.2in}{2.7in}{\includegraphics{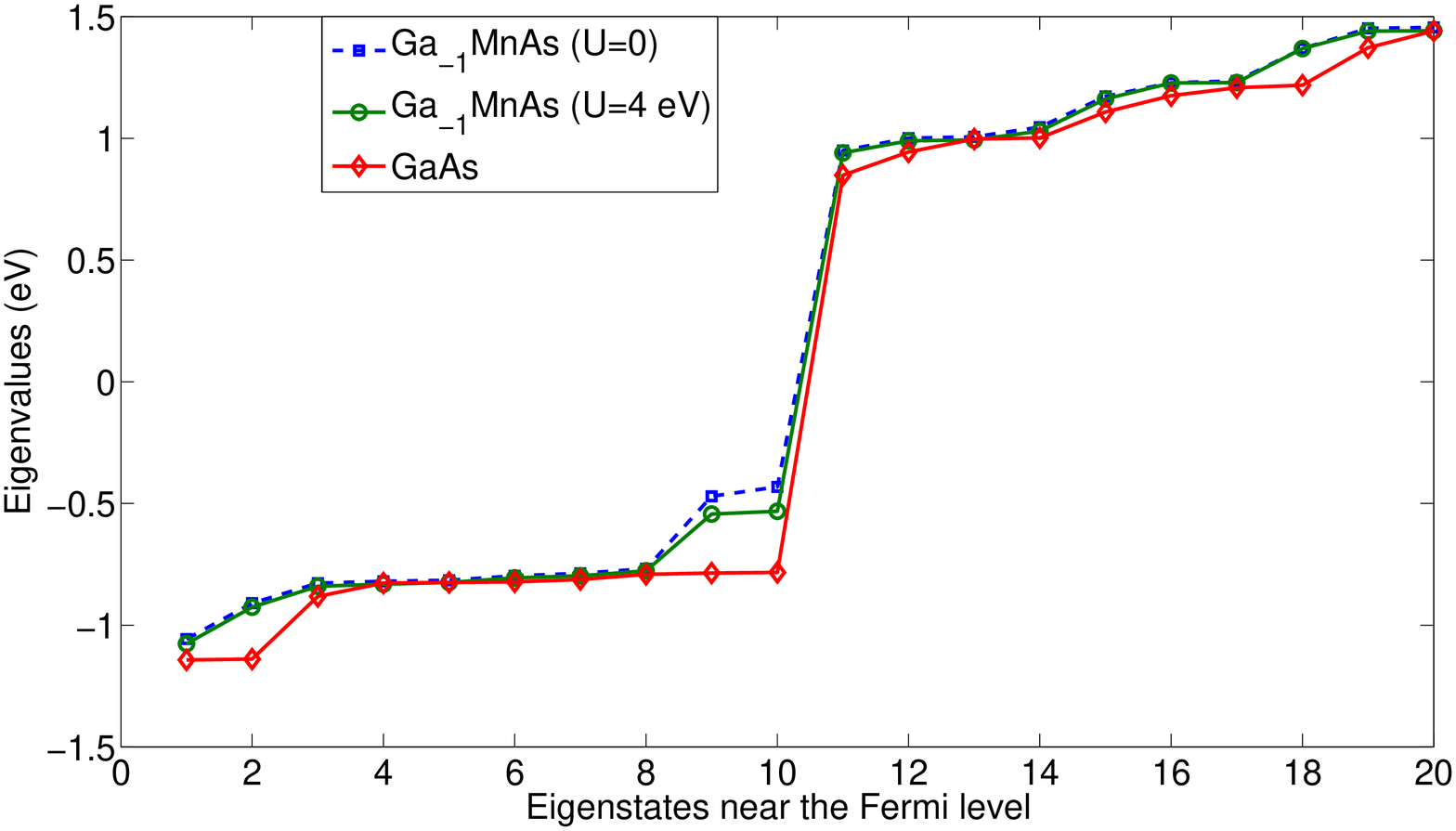}}}
\caption{(Color online) Eigenvalue energies of eigenstates in the
surroundings of the GaAs energy gap. The red (gray line) is for pure
GaAs. The blue (solid black) and green (dashed) curves are for the
case when a surface Mn impurity is present (solid black curve) or
not (dashed) green. Note the appearance of the hole acceptor state
in the gap, a few hundred meV above the GaAs valence-band edge. The
acceptor state energy is moderately affected by the presence of
Hubbard $U$ correlations, a sign that this state is in part
hybridized with d-states.} \label{eigenvalues}
\end{figure}

It is useful to look more carefully at the effect of the Hubbard
correlations on the properties of the hole state. In
Fig.~\ref{eigenvalues} we plot the discrete eigenvalue energies
\footnote{In our calculations we employ one $k$-point only. The
discrete spectrum is the solution of the Kohn-Sham equations for
this $k$-point only.} for the states in the surroundings the GaAs
energy gap, for pure GaAs  (red curve) and for the case in which a
Mn impurity is present (blue and green curves). As a result of the
Mn presence, acceptor states appear in the gap of the host material.
Specifically, we find two energy levels located a few hundred meV
above the GaAs valence-band edge. We identify the topmost of these
two levels as the acceptor state, occupied by a hole. The energy
immediately below corresponds to the highest occupied (by an
electron) level. Note that, with the exception of one level deep in
the valence-band, the emergence of these two mid-gap levels is the
only important qualitative difference that distinguishes the
spectrum of the system in the presence of the Mn impurity from the
one of pure GaAs. As shown in the figure, on site $U$ correlations
have an effect on these two impurity levels, slightly decreasing
their energy toward the valence band-edge. The sensitivity of the
acceptor energies to the Hubbard $U$ parameter, indicates that these
levels are indeed moderately hybridized with the Mn d orbitals.

In order to understand further the effect of correlations on the
hole state, in Tables~\ref{occu0} and \ref{occu4} we display the
square of the wave function for a few levels around the Fermi level
at the Mn site and at the site of its two As nearest neighbors (nn)
on the (110) surface.  The partial contribution of a given orbital
character ($p$ or $d$) is shown together with their sum for a given
atom. Table~\ref{occu0} and \ref{occu4} correspond to the cases of
$U =0$ and $U=4$ respectively. As anticipated above, we can see that
the hole state comes about from the hybridization of the Mn d
orbitals with the $p$ states of its two As nn. We note that with
$U=0$  (see Table ~\ref{occu0}) the hole state is predominantly of
$d$ character but hybridized with As $p$ states. With $U$ turned on
(see Table ~\ref{occu4}), the $d$ character of the hole state at the
Mn site decreases, while the $p$ character weight on the nearby As
is essentially unchanged. This is a clear indication that the hole
state is pushed out into the interstitial region between Mn and its
nearby As atoms by on-site correlations, namely it acquires a less
localized and more itinerant character.

\begin{table}
\begin{tabular}{|c|c|c|c|c|}   \hline
Energy   & Atom & \multicolumn{3}{c|}{$|\psi_{{\rm atom},\mu}|^2 $}  \\ \cline{3-5}
(eV)     &      &  Total  &  p    &    d      \\  \hline
         & Mn   & 0.002 & 0.000   & 0.002     \\  \cline{2-5}
-0.7681  & As1  & 0.004 & 0.004   & 0.000     \\  \cline{2-5}
(valence)         & As2  & 0.004 & 0.004   & 0.000     \\  \hline
         &  Mn   &  0.102   &  0.029  &  0.073     \\  \cline{2-5}
 -0.4700  &  As1  &  0.101 &  0.100 &  0.001
\\  \cline{2-5}
(highest occ.)
         &  As2  &  0.099 &  0.098 &  0.001     \\  \hline
         &\bf  Mn   &\bf  0.116 &\bf  0.017  &\bf  0.095      \\ \cline{2-5}
\bf -0.4321  &\bf  As1  &\bf  0.069 &\bf  0.065  &\bf  0.000  \\ \cline{2-5}
\bf  (hole)  &\bf  As2  &\bf  0.070 &\bf  0.066  &\bf  0.001  \\ \hline
         & Mn   & 0.001  & 0.000  & 0.001      \\  \cline{2-5}
 0.9500 & As1  & 0.004  & 0.004  & 0.000      \\  \cline{2-5}
(conduction)   & As2  & 0.004  & 0.004  & 0.000      \\  \hline
\end{tabular}
\caption{Square of the wave function, $|\psi_{{\rm atom},\mu}|^2 $,
of different energy levels near $E_F$, including the acceptor hole
state, calculated at the Mn site and its two As nn on
the (110) surface. The $\mu = p, d$ orbital character contributions
are separately specified on the second and third column. Here the
Hubbard $U=0$. The first level belongs to the host valence band,
while the last level is in the conduction band. The bold faced
numbers correspond to the acceptor hole state. The level immediately
below the hole state corresponds to the highest occupied level. Both
states are inside the energy gap. See Fig.~\ref{eigenvalues}.}
\label{occu0}
\end{table}

\begin{table}
\begin{tabular}{|c|c|c|c|c|}   \hline
 Energy  & Atom & \multicolumn{3}{c|}{$|\psi_{{\rm atom},\mu}|^2 $}  \\ \cline{3-5}
 (eV)    &      &  Total  &   p    &    d       \\  \hline
         & Mn   & 0.001   & 0.000  & 0.001      \\  \cline{2-5}
-0.7764  & As1  & 0.005   & 0.005  & 0.000      \\  \cline{2-5}
(valence)         & As2  & 0.005   & 0.005  & 0.000      \\  \hline
         &  Mn   & 0.052   & 0.029  & 0.023      \\  \cline{2-5}
 -0.5431 & As1  & 0.107   & 0.107  & 0.000      \\  \cline{2-5}
(highest occ.)         & As2  & 0.103   & 0.103  & 0.000      \\  \hline
         & \bf Mn   & \bf 0.052  &\bf  0.016 &\bf  0.032     \\  \cline{2-5}
\bf -0.5316  & \bf As1  &\bf  0.069  & \bf 0.065  & \bf 0.000 \\  \cline{2-5}
\bf (hole)  & \bf As2  &\bf  0.073  & \bf 0.069 & \bf 0.000      \\  \hline
         & Mn   & 0.001  & 0.000  & 0.000      \\  \cline{2-5}
 0.9400 & As1  & 0.004  & 0.002  & 0.000      \\  \cline{2-5}
(conduction)   & As2  & 0.004  & 0.004  & 0.000      \\  \hline
\end{tabular}
\caption{Square of the wave function, $|\psi_{{\rm atom},\mu}|^2 $,
as in Table \ref{occu0}, but with $U= 4$eV.} \label{occu4}
\end{table}

As shown by Mahadevan {\it et al.},\cite{zunger2} on-site
correlations are known to increase the itinerant character of the
hole state for Mn impurities in bulk GaAs\cite{zunger2}. To
understand the difference in the nature of the hole state in bulk
and surface, we have calculated the Mn acceptor hole properties of
bulk GaAs. We use a 64 atom supercell with a single Mn impurity in
the middle. The occupancies of the hole state at the Mn site, its
two nn and two next nearest neighbor (nnn) As sites are tabulated in
Table~\ref{bulk_occu}. It is evident that when $U$ is turned on the
hole state extends over nnn As sites. A similar result is obtained
by Mahadevan {\it et al.}\cite{zunger2} in their bulk GaAs
calculations. In the surface calculations, although the occupancies
of the hole state at the nnn As atoms on the surface layer slightly
increases, the effect is less pronounced in the surface compared to
the bulk. We conclude that for a Mn impurity on the surface, the
associated acceptor state remains predominantly localized in the
interstitial region between the Mn and the two nn As atoms.
Consequently, the hole state for a Mn on the (110) is energetically
a rather deep impurity, compared to the case of Mn in bulk GaAs. The
localized character of the hole wavefunction for a surface Mn and
its relatively large binding energy are supported by both STM
experiments\cite{yazdani06,garleff_prb_2010, gupta_science_2010} and
tight-binding calculations.\cite{canali09}

Finally, Tables  \ref{occu0} and \ref{occu4} show that on-site
correlations also decrease the Mn $d$-orbital contribution of the
highest occupied level by approximately 50\% (compare Table
\ref{occu0} and, \ref{occu4}). As evident from the inset of
Fig.~\ref{d_dos}, this effect involves primarily majority spin
states at the Fermi level.

\begin{figure}[htbp]
{\resizebox{3.2in}{2.2in}{\includegraphics{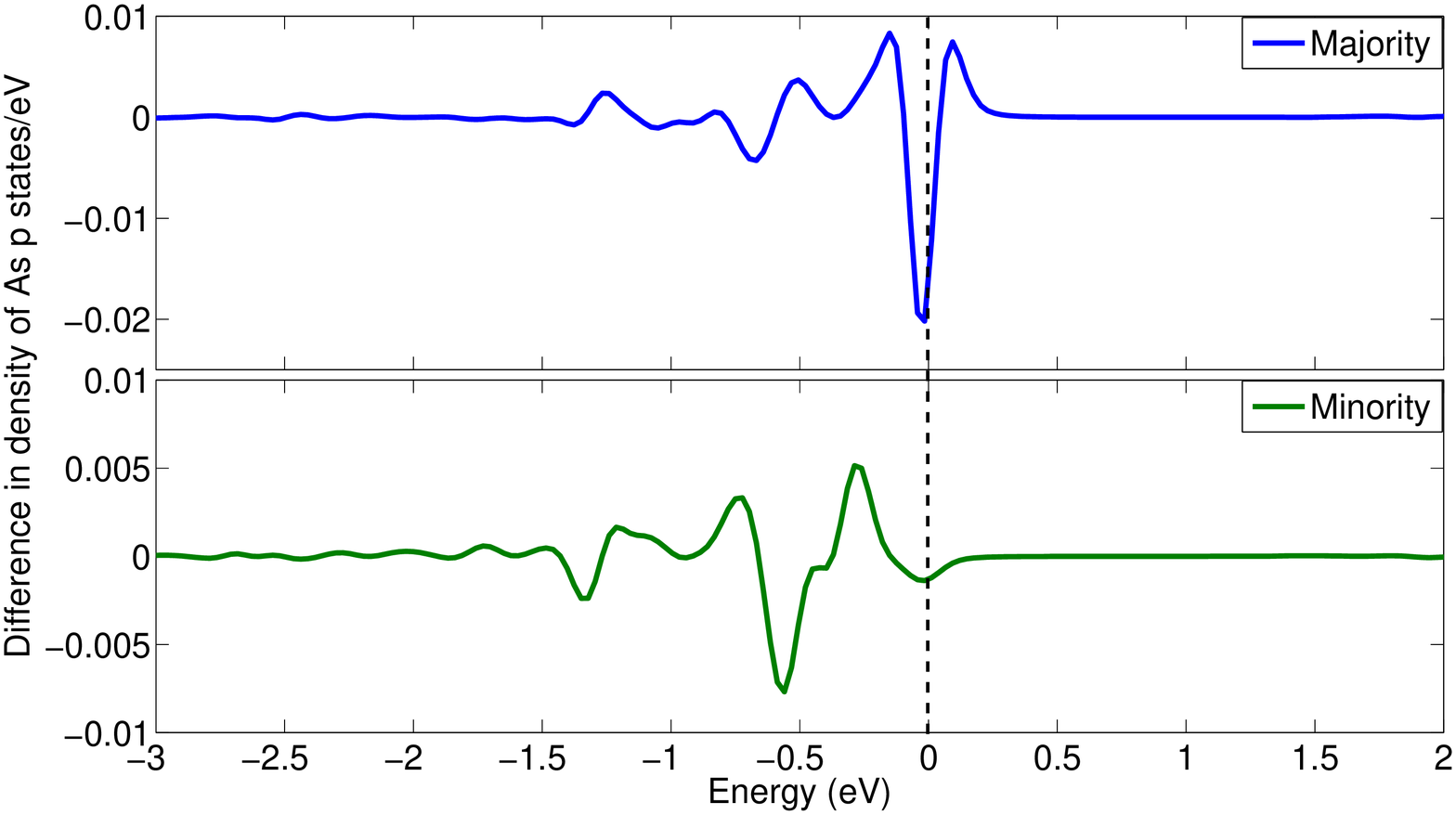}}}
\caption{(Color online) Difference in the As $p$ density states, for
two directions of the magnetization, $<001>$ and $<010>$, in the
presence of SOI and for $U=4$ eV.} \label{diff_p_dos}
\end{figure}

The electronic properties the Mn acceptor hole state are only weakly
affected by the direction of the Mn moment direction as a result the
SOI. In Fig.~\ref{diff_p_dos} we plot the difference in the $p$
partial DOS for one of the two As nearest-neighbor, for two
directions of the magnetization, $<001>$ and $<010>$ respectively.
The SOI induced change is only a few percents. We will see below
that the magnetic anisotropy energy can be explained by this
direction dependence. This is consistent with tight binding
calculations\cite{canali09} for a substitutional Mn on the (110)
surface, which show that the square of acceptor hole wave function
is anisotropic in space, but does not change significantly with the
Mn moment direction.

\begin{table}
\begin{tabular}{|c|c|c|c|c|c|c|}   \hline
Atom  & \multicolumn{6}{c|}{$|\psi_{{\rm atom},\mu}|^2 $}  \\ \cline{2-7}
      & \multicolumn{3}{c|}{$U$=0} & \multicolumn{3}{c|}{$U$=4 eV} \\  \cline{2-7}
      & Total &   p   &   d   & Total &   p   &   d    \\   \hline
Mn    & 0.176 & 0.024 & 0.151 & 0.050 & 0.013 & 0.036  \\   \hline
As1+As2  & 0.098 & 0.092 & 0.002 & 0.078 & 0.076 & 0.002 \\
(nn)     &       &       &       &       &       &       \\ \hline
As3+As4  & 0.006 & 0.006 & 0.000 & 0.018 & 0.018 & 0.000   \\
(nnn)    &       &       &       &       &       &       \\   \hline
\end{tabular}
\caption{Properties of the hole state, as in Tables \ref{occu0},
\ref{occu4}, but for a Mn in {\it bulk} GaAs on the (110) plane. As3
and As4 are the two next nn As atoms of the Mn on the (110) plane.}
\label{bulk_occu}
\end{table}

\subsubsection{Spin magnetic moment}
The spin magnetic moment $M_S$ for one Mn on the (110) GaAs surface
is given in Table \ref{spin_mom}. When $U=0$  the spin moment on the
Mn is $M_S = 4.05 $ in units of the Bohr magneton $\mu_{\rm B}$. A
close inspection shows that there are 4.39 majority-spin and 0.42
minority-spin $d$ electrons on the Mn atoms. The remaining small
contribution 0.08 to the Mn spin moment comes from $s$ electrons.
The deviation from the Mn atomic limit of 5 $\mu_B$ is due to the
acceptor hole being in part localized on the Mn. When $U =4 $, the
spin on the Mn increases from 4.05 to 4.46 due to the concomitant
increase of the majority-spin and decrease of the minority-spin $d$
electrons. The increase of the spin moment on the $3d$ adatom as a
result of correlations is expected, since the Hubbard $U$ term tends
to localize the $d$ electrons. Notice however that, in this case,
the total magnetic moment per unit cell $M_S^{\rm Tot} = 4.21 $ is
smaller that $M_S$. Indeed $M_S^{\rm Tot}$ includes also a negative
contribution from the $p$ orbitals on the nearby As atoms, which are
partly spin polarized in the opposite direction of the Mn $d$ spin
moment, as shown in Fig.~\ref{p_dos}. The results in Table
\ref{spin_mom} are very weakly affected by SOI. In particular, the
value of the spin magnetic moment does not show any appreciable
anisotropy with respect to its direction.

\begin{table}[ptb]%
\begin{tabular}
[c]{ccc}\hline\hline
{Spin moment, $\mu_{\rm B}$} & $M_S$ & $M_S^{\rm Tot}$ \\\hline
$U=0$   & 4.05& 4.07\\
$U=4$   & 4.46& 4.21\\
\hline\hline
\end{tabular}
\caption{Spin magnetic moment on Mn ($M_S$) and total spin moment
per unit cell ($M_S^{\rm Tot}$) in Bohr magnetons.} \label{spin_mom}
\end{table}

\subsubsection{Orbital moment}

Next we discuss the orbital magnetic moment. An isolated Mn atom has
the spin-majority d levels fully occupied and the spin-minority
states empty. As a result the total orbital moment $M_L$  is zero.
Note that each of the individual degenerate $d$ orbitals carries a
unit of orbital angular momentum equal to  $m_l = 0 , \pm 1, \pm 2$.
When a Mn atom is substituted for a Ga atom in GaAs, the crystal
field of the host material removes the degeneracy of the Mn $d$
levels and each orbital moment of the perturbed states becomes
quenched. The effect of SOI is to reduce quenching and to restore,
at least partially, the individual orbital moments by mixing
different $d$ states. However, since the spin-majority $d$ manifold
is still practically fully occupied while the minority-spin is
empty, we expect that, even in the presence of SOI, the total
orbital moment $M_L$ on the Mn atom to be very small. On the other
hand, an acceptor Mn impurity introduces also an itinerant hole
state, which is a state with a predominant $p$ character and as such
should have an orbital magnetic moment. Thus we expect that the
valence band states, filled up to the Fermi level, should generate a
net orbital moment equal and opposite to the orbital moment of the
hole. These simple considerations are fully supported by the DFT
calculations. Results are summarized in Table \ref{orbit_mom}.

\begin{table}
\begin{tabular}{|c|c|c|c|c|}   \hline
Magnetization  & \multicolumn{4}{c|}{Orbital moment, $\mu_B$}  \\ \cline{2-5}
direction      & \multicolumn{2}{c|}{$U$=0} & \multicolumn{2}{c|}{$U$=4 eV}      \\  \cline{2-5}
               & $M_L^d$ & $M_L^{p, {\rm Tot}}$ & $M_L^d$ & $M_L^{p, {\rm Tot}}$ \\  \hline
$\langle 001\rangle $ & 0.015 & -0.028 & 0.007 & -0.031      \\ \hline
$\langle 010\rangle $ & 0.007 & -0.016 & 0.000 & -0.019   \\ \hline
\end{tabular}
\caption{Orbital $d$ magnetic moment on Mn ($M_L^d$) and total $p$
orbital moment per unit cell ($M_L^{\rm Tot}$) in Bohr magnetons
calculated for two directions of the magnetization, $\langle
001\rangle$ (easy) and $\langle 010\rangle$ (hard). The minus sign
indicates that the orbital moment points in the opposite direction
of the spin orbital moment.} \label{orbit_mom}
\end{table}

We find that the calculated orbital moment of the Mn $d$ states is
indeed very small and slightly anisotropic: for the magnetization
directed along $\langle 001\rangle$ (which turns out to be the easy
direction, see next Section) $M_L^d \approx 0.015 \mu_B$; for the
magnetization directed along $\langle 010\rangle$ (one of the hard
axis), $M_L^d$ is about 0.007 $\mu_B$. The total $p$ orbital moments
of the unit cell for easy and hard directions are -0.028 and -0.016
$\mu_B$, respectively. These value are larger than the $d$ orbital
moment but still considerably small. The negative moment implies
that orbital moment is opposite to the spin moment of the system.
The orbital moment of the hole state, on the other hand, is aligned
parallel to the spin moment and, as expected form the discussion
above, almost exactly cancels the orbital moment of the whole cell.
The largest contribution to the moment comes from the two As (14\%
each) nearest to Mn (5\%) and one As in the first subsurface layer
(12\%). For the easy axis, apart from these atoms other subsurface
layer As atoms along $[111]$ direction also have finite contribution
to the orbital moment, suggesting that the hole state is extended
along that direction. For the hard axis the direction of extension
of the hole state remains the same but the largest contribution to
the $p$ orbital moment comes from the first subsurface layer As atom
(16.5\%), while two surface As atoms contribute only 7.4\% each with
negligible contribution from Mn. The on-site correlation, $U$, has
very little effect on p orbital moment of the whole cell but the d
orbital moment of Mn becomes vanishingly small with $U=4$ eV.

On the basis of these results for the spin and orbital magnetic
moment, we can conclude that for a Mn impurity on the (110) surface,
the total magnetic (spin plus orbital) moment of the system is
approximately 4 $\mu_{\rm B}$. Assuming that the total magnetic
moment is a combined effect of the Mn and the acceptor hole, we can
imagine to assign an effective "spin" $J= 2$ to this magnetic entity
composed of the Mn impurity and its hole. The effective spin $J$ can
be understood as a result of the Mn spin ($S= 5/2$)
antiferromagnetically coupled to the spin of the hole state
($s=1/2$). For both Mn and acceptor the associated orbital moment is
small, although not entirely negligible for the hole. These results
can be compared with the well-known situation of a substitutional Mn
impurity in bulk GaAs. In that case electron-spin
resonance\cite{schneider_prb87} and infrared spectroscopy absorption
experiments,\cite{linnarsson_prb97} together with theoretical
considerations, have shown that the the total spin of the (Mn +
hole) complex is $J= 1$, and it results from the antiferromagnetic
coupling of the Mn spin moment $S=5/2$ and the hole total angular
momentum $j= s+ l= 1/2 + 1= 3/2$. Our calculations indicate that
when a Mn replaces a Ga on the Mn surface, the orbital moment of the
associated acceptor hole is small, most likely due to the fact that
the high (tetragonal) symmetry experienced by the hole in the bulk
GaAs is strongly reduced by the surface. This reasoning and
conclusions are in agreement with a recent theory\cite{canali11}
that identifies the total effective "spin" of the (Mn + hole)
acceptor magnet with a Berry phase Chern number $J$. Tight-binding
calculations implementing this theoretical approach find that $J$ is
2 for a Mn near the (110) surface, due to the quenching of the
orbital moment of the acceptor hole state.

\subsubsection{Magneto-crystalline anisotropy}
\label{mae} We now discuss the magnetic anisotropy energy (MAE),
defined as the dependence of the ground-state energy on the
magnetization direction. The most important contribution  to the MAE
is the magneto-crystalline anisotropy caused by the SOI. For a
single substitutional Mn impurity in bulk GaAs the tetragonal
symmetry of the host lattice implies that the MAE is essentially
zero. On the surface however, the symmetry is broken, and the
anisotropy is finite. Note that surface relaxation and strain (e,g.
induced by interfaces with other lattices or by external electric
fields) can enhance the anisotropy considerably.

Although it is common to calculate the MAE using an approximate
method known as magnetic force theorem,\cite{lichtenstein_jmmm_1987}
we have first estimated it by calculating the total energy including
SOI as a function of different magnetization directions shown in
Fig.~\ref{surf_relx}. The anisotropy barrier between two
magnetization directions, $\hat M_S$ and $\hat M_S'$, is defined as
the difference in total energy between these two directions,
\begin{equation}
\Delta E_A = E(\hat M_S)-E(\hat M_S')
\label{anisotropy}
\end{equation}

In our calculations we find that the total energy $ E(\hat M_S)$ is
minimal when $\hat M_S$ is along the $\langle 001\rangle $ direction
(easy axis) and maximal when $\hat M_S$ is along the $\langle
010\rangle $ direction (hard axis), both of which lie in the (110)
plane. The largest MAE barrier that we find from our calculation is
1.17 meV when $U=0$ and it decreases slightly to 1.13 meV when $U=
4$ meV. A summary of the magnetic anisotropy energy for different
magnetization orientations is presented in Table~\ref{MAE}. Notice
that the MAE barrier to rotate the magnetic moment from the easy
axis and align it along the $<100>$ direction (perpendicular to the
surface) is only 0.16 meV.

\begin{table}[ptb]%
\begin{tabular}
[c]{ccc}\hline\hline
{MAE} & $U=0$ & $\ U= 4$ eV \\\hline
$E(\langle 010\rangle) - E(\langle 001\rangle)$   & 1.17 & 1.13 \\
$E(\langle 100\rangle) - E(\langle 001\rangle)$   & 0.17 & 0.16 \\
$E(\langle 010\rangle) - E(\langle 100\rangle)$   & 1.00 & 0.97 \\
\hline\hline
\end{tabular}
\caption{Magnetic anisotropy energy (in meV) for one substitutional
Mn impurity on the (110).}
\label{MAE}
\end{table}

It is now instructive to look at the MAE form the point of view of
the magnetic force theorem.\cite{lichtenstein_jmmm_1987} This
amounts to express the total energy $E(\hat M_S)$  as the sum of the
magnetization-dependent Kohn-Sham band eigenvalues, $\epsilon_n$

\begin{equation}
E(\hat M_S) = \sum_n^{\rm occ} \epsilon_n (\hat M_S)
\label{magnetic_force}
\end{equation}

where the sum is over the occupied KS eigenvalues. By introducing
the (magnetization dependent) DOS, we can rewrite $E(\hat M_S)$ as

\begin{equation}
E(\hat M_S) = \int_{\epsilon_{\rm B}}^{\epsilon_{\rm F}}
(\epsilon -\epsilon_{\rm F})\rho(\hat M_S)  d \epsilon
\label{magnetic_force2}\;,
\end{equation}

where ${\epsilon_{\rm B}}$ is the bottom of the valence band and
${\epsilon_{\rm F}}$ is the Fermi energy. We can then write the MAE
barrier as

\begin{equation}
{\rm MAE}= \int_{\epsilon_{\rm B}}^{\epsilon_{\rm F}}
(\epsilon -\epsilon_{\rm F})
[\rho(\hat M^{\rm hard}_S) - \rho(\hat M^{\rm easy}_S) ]  d \epsilon
\label{MAE_m_force}\;.
\end{equation}

It turns out that the integrand of Eq.~\ref{MAE_m_force} -- the
change in band energy upon varying the direction of the
magnetization -- can be of either sign. When these changes are
summed up for all occupied states we can expect large cancellations,
so that what matters in the end is the dependence of the band
energies around the Fermi energy.

We can now surmise that the magnetic properties of the system
including the variations of the DOS on the magnetization direction,
should mainly affect the region of the Mn and its immediate
surroundings. As we can see in Fig.~\ref{d_dos} and
Fig.~\ref{p_dos}, the projected DOS at the Fermi energy for the Mn
impurity and its nearest-neighbor As atoms shows that the dominant
contribution come from the states of $p$ character primarily located
on the nn As. An inspection of the $p$ DOS for many other atoms in
the supercell shows that, besides the Mn and its two nn As atoms,
only the As atoms in the first sub-layers closest to the Mn have a
finite contribution at $E_{\rm F}$, which becomes progressively
smaller the further the As is from the Mn. Thus we can expect that
the MAE is primarily controlled by the $p$ orbital components of the
band energies at $E_{\rm F}$ located at these atoms.

Indeed, if we calculate the MAE by integrating Eq.~\ref{MAE_m_force}
for all band energies within a few eVs of the Fermi level for the Mn
and the nearby As atoms, including those As on the sub-surface
layers, we obtain ${\rm MAE} \approx 2.8 $ meV, which differs only
by a factor of 2 from the more precise estimate obtained above. The
two As atoms nearest to Mn on the surface layer contribute about
$64\%$ of the MAE, whereas the As on the first sub-surface layer and
closest to Mn contribute about $17\%$, with Mn $d$ level
contributing about $18\%$.

Furthermore, tight-binding calculations show that the MAE obtained
by summing up magnetization-dependent energy variations for all
occupied levels is equal and opposite of the anisotropy energy for
the acceptor hole state only\cite{canali09}. If we compute the
integral $\rm MAE_{hole}= \int_{\epsilon_{\rm F}}^{\epsilon_{\rm
hole}} (\epsilon -\epsilon_{\rm F}) [\rho(\hat M^{\rm hard}_S) -
\rho(\hat M^{\rm easy}_S)] d \epsilon$, where $\epsilon_{\rm hole}$
extends up to the limit of the hole-peak DOS (see Fig.~\ref{p_dos}),
we find $\rm MAE_{hole}\approx -2.25$ meV, which has the expected
opposite sign and is very close in magnitude to the MAE calculated
summing all the energy shifts up to $E_{\rm F}$. About $68\%$ of the
hole anisotropy comes from the two As atoms nearest to Mn on the
surface layer, while the Mn and the first sub-surface layer As
contributions are $14\%$ and $16\%$ respectively.

Note that in the previous section we found that a similar property
holds for the the orbital moment of the system: the total orbital
moment, primarily located on the Mn, the nn As, and the closest As
in the immediate subsurface layers, is equal and opposite of the
contribution of the hole-state orbital moment, located primarily on
the same atoms.

The difference in MAE calculated from the total energy difference
(Eq.~\ref{anisotropy}) and magnetic-force theorem
(Eq.~\ref{MAE_m_force}) is expected, since the latter is an
approximate method, and is due to the fact that DFT calculates the
groundstate energy of a many-body system in a physically meaningful
way, whereas individual single-particle energies are rather a
mathematical artifact used to solve the Kohn-Sham equations.

In applying the magnetic-force theorem, we have disregarded the
contribution of all the other atoms in the supercell, under the
assumption that their contributions are small around the Fermi
level. In fact, we have explicitly verified that the $p$ partial DOS
of these As atoms have a negligible value around the Fermi level and
the integration of Eq.~\ref{MAE_m_force} around the Fermi level
gives insignificant contribution to the MAE. This is, however, not
the case if we include the band energies deep in the valence band in
the calculation of the MAE, which appear to give some contribution
for all atoms. We don't believe that this is an indication of a
finite contribution to the MAE coming from atoms far way from the
Mn, but rather a sign of the non-sufficient accuracy in the
computation of the SOI-induced shift in the states deep in the
valence band.

At this point it is interesting to investigate more closely the
connection between the anisotropy properties of the total energy and
the orbital moment. The discussion above already suggests that these
two quantities should be related. According to the perturbative (in
SOI) analysis by P. Bruno,\cite{Bruno89} there is an approximate
relationship between the MAE and the orbital moment anisotropy
\begin{equation}
E(\hat M_S) - E(\hat M_S') = -\frac{\xi}{4}\big[ M_L(\hat M_S) - M_L(\hat M_S')\big]
\label{bruno}
\end{equation}
where $\xi$ is the SOI coupling strength. In particular,
Eq.~\ref{bruno} implies that $\Delta E_A$ is proportional to the
difference in the orbital moment for the spin moment in the easy and
hard direction. The orbital moment calculation clearly shows that
the orbital moment anisotropy of the hole comes mainly from As $p$
states with a small contribution from the Mn $d$ states. Using the
SOI coupling constant of As, $\lambda_{As}$=140 meV,\cite{chadi78}
Bruno formula yields $\Delta E_A^{\rm Bruno}$=0.35 meV. This is
approximately a factor of 3 smaller than MAE calculated using DFT.
The Bruno formula has been investigated recently for systems
consisting of magnetic impurities on metal\cite{blonski_prb_2010}
and insulator surfaces.\cite{lichtenstein_prb_2009} In some cases
the relationship seems to work satisfactorily, in other it fails. As
far as we know a similar analysis for magnetic acceptor impurities
in semiconductors has not been carried out. This situation is
different from an ordinary magnetic impurity on an insulator for
example, in that, an important role in the magnetic properties of
the system is now played by the itinerant acceptor hole. Our
analysis shows that orbital moment of the hole state is certainly
related to the magnetic anisotropy energy. as in Eq.~\ref{bruno},
albeit with a renormalized SOI coupling strength. This might be an
indication that our DFT calculations underestimate the orbital
magnetic moment due to the lack of orbital dependence of the
exchange potential.\cite{blonski_prb_2010}

The electronic and magnetic properties of a single Mn impurity on
the (110) GaAs surface calculated here are in general, in good
agreement with tight-binding based calculations by Strandberg {\it
et al.}\cite{canali09} In particular, the value of the magnetic
anisotropy energy barrier are pretty close for the two approaches
$\approx $ 1 meV. However, details of the magnetic anisotropy energy
landscape are different. While in our calculation we have both easy
and hard axes in the (110) plane, Strandberg {\it et al.} find an
easy axis that makes an angle $45^0$ with the surface. The
difference may stem from the fact that they have used much larger
clusters and hence their system is much more diluted than the system
used in this calculation.

\subsection{Mn pair on (110) GaAs surface}
\label{results_section_2Mn}

In this section we discuss the properties of a pair of
substitutional Mn impurities on the $(110)$ surface. It turns out
that the electronic and magnetic properties are quite similar to the
ones of individual Mn impurities. After a brief summary of these
properties we will concentrate on discussing the effective exchange
coupling and the magnetic anisotropy of the Mn pair and their
dependence on atom separation and orientation on the $(110)$
surface.

\subsubsection{Electronic structure and spin moment}

\begin{figure}[htbp]
{\resizebox{3.2in}{2.2in}{\includegraphics{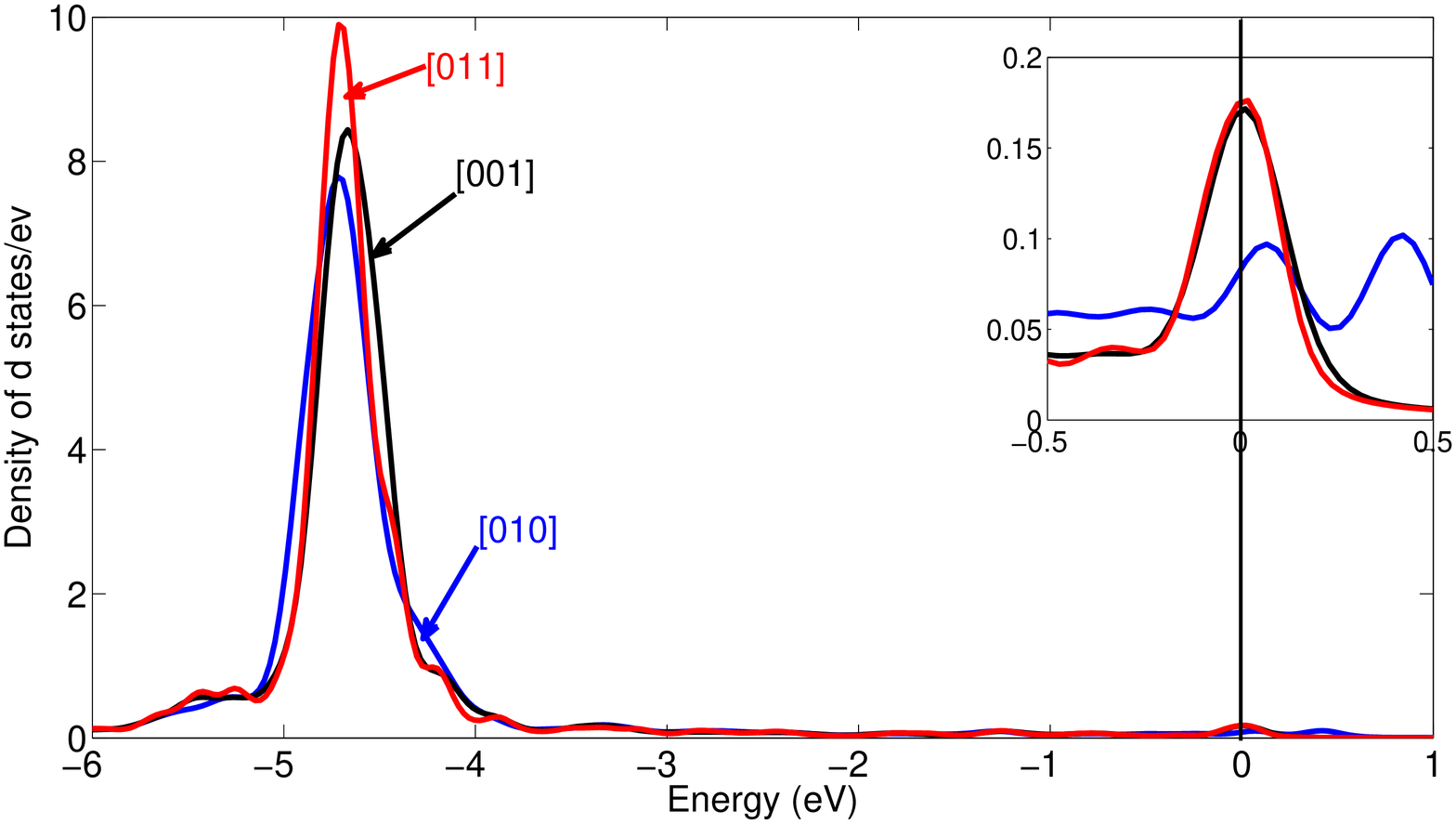}}}
\caption{(Color online) Density of $d$ states for different
orientations of a pair of Mn on (110) surface of GaAs. See
Fig.~\ref{surf_relx} for different orientations of the pair relative
to the crystallographic axes. The inset shows the DOS near the Fermi
level.} \label{surf_dos}
\end{figure}

The $d$ partial DOS for different pairs of Mn impurities
shows a similar pattern as in the case of a single impurity, both
with $U=0$ and $U=4$ eV. In particular there is rather weak
dependence on atom separation and pair orientation with respect to
the surface crystal structure. As for the case of individual Mn
impurities, the effect of $U$ is to reduce the intensity near the
Fermi level and to further split the main majority and and minority
peaks, pushing the former below the Fermi level and the latter
above. Fig.~\ref{surf_dos} shows the local density of $d$ states of
one of the two Mn atoms (up spin only) for  different pairs on the
surface, when $U$=4 eV. While the main majority peak is insensitive
to the orientation of the Mn pairs, the partial DOS is rather wide at the
Fermi level for the [010] orientation because of the smaller
distance between the two Mn atoms for this orientation.

The properties of the spin magnetic moment (per atom) are
essentially the same as for individual Mn impurities, for all pairs
on the surface. For example, the Mn spin moment is approximately
equal 4 $\mu_B$ when $U=0$ and increases to 4.5 $\mu_B$ when on-site
correlations are included by setting $U=4$ eV, as in the case of a
single impurity.

\subsubsection{Exchange interaction}

A strong exchange interaction between magnetic impurities in a
semiconductor is essential for its successful application as a
spintronic device, since magnetic properties of semiconductors must
be preserved at room temperature and above. In this work we have
calculated the exchange energy between two Mn impurities on (110)
surface of GaAs and have investigated the effect of correlations and
SOI on exchange. The exchange energy for each pair orientation is
defined as the difference in energies of the supercell when the
spins of the two impurity atoms are arranged parallel and
antiparallel

\begin{equation}
J\equiv E_{\rm cell}^{\uparrow\downarrow}- E_{\rm cell}^{\uparrow\uparrow}
\end{equation}

\begin{figure}[htbp]
{\resizebox{3.3in}{2.2in}{\includegraphics{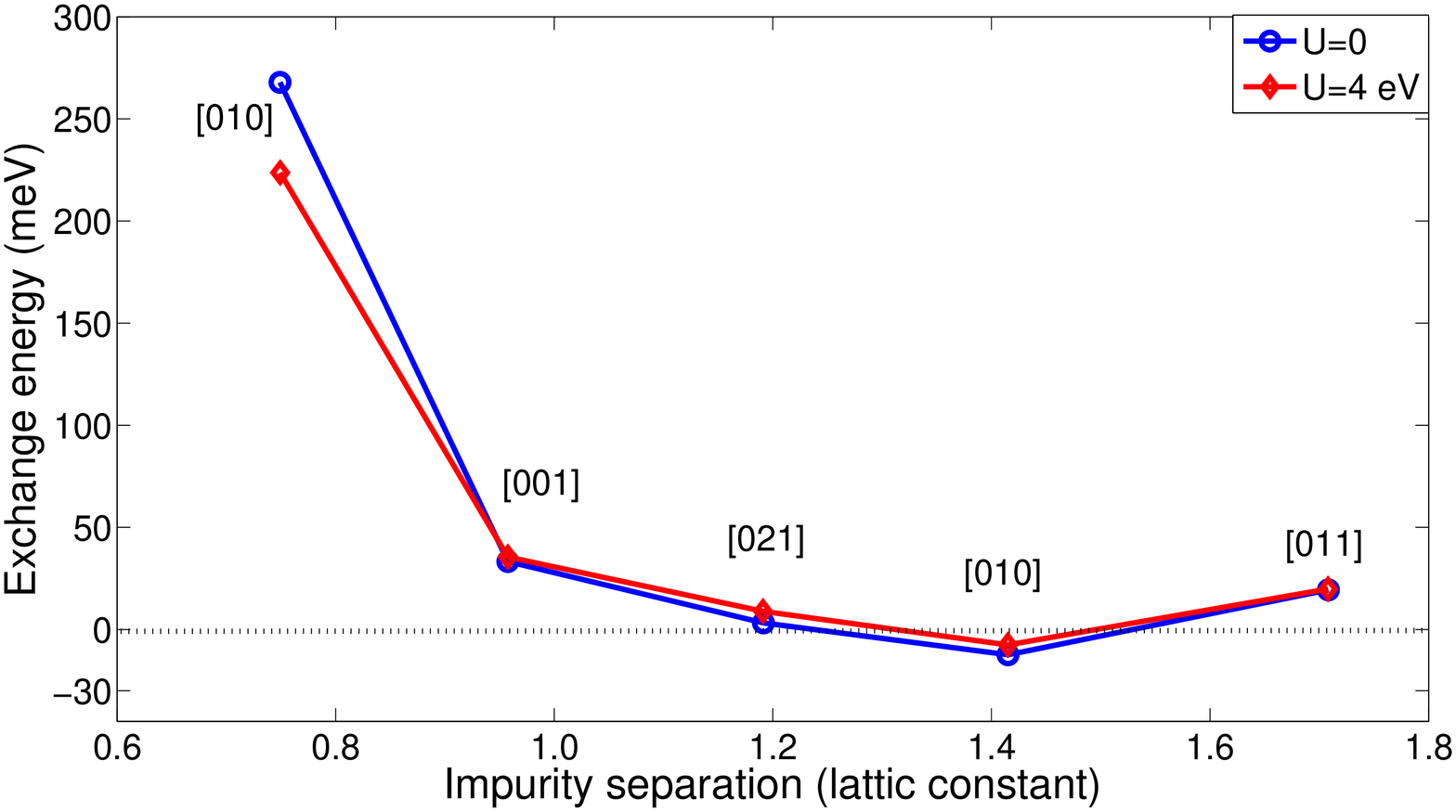}}}
\caption{(Color online) Dependence of the exchange energy constant
$J$ on the orientation and Mn-separation for a pair of Mn impurities
on the (110) surface of GaAs.} \label{surf_Mn}
\end{figure}

A positive (negative) $J$ implies that the Mn-Mn interactions are
ferromagnetic (antiferromagnetic) Fig.~\ref{surf_Mn} shows the
dependence of the exchange constant $J$ on the separation between
the two Mn atoms of the pair and on the pair orientation relative to
the crystallographic axes of the surface.

The exchange constant $J$ is largest for the pair with the shortest
separation along the $[010]$ direction, where the two Mn are nearest
neighbors, and it decays quickly with impurity separation. The
general trend and order of magnitude of $J$ shown in
Fig.~\ref{surf_Mn} is in agreement with the results of tight-binding
calculations, despite these are carried out for a much more dilute
system, which is supposed to model the situation studied in STM
experiments. Although the smaller size of the surface supercell
restricts our ability to study, in detail, the orientation
dependence of exchange, it is evident from the figure that the
exchange constant is larger when the pair is oriented along [011]
direction than when it is oriented along [021] direction, even
though the distance between the Mn atoms is smaller in latter case,
This is a clear indication of the strongly anisotropic nature of the
exchange coupling for pairs of substitutional Mn impurities on the
(110) GaAs surface. Evidence of this anisotropy was found in recent
STM experiments\cite{yazdani06} and is also supported by
tight-binding
calculations.\cite{tangflatte_prl04,yazdani06,tangflatte_spie_2009,
canali10} We also note that as the distance between Mn atoms
increases along the [010] direction, $J$ becomes negative, i.e. the
antiferromagnetic coupling of the Mn pair becomes energetically
favorable. This result implies that a random distribution of
impurities along different directions might reduce the overall
magnetization of a dilute system.

\begin{table}[h]
\label{tablesurf}
\begin{tabular}{||c|c|c|c|c||}   \hline
Orientation   & \multicolumn{2}{c|}{Exchange} & \multicolumn{2}{c|}{Magnetic} \\
of Mn atoms   & \multicolumn{2}{c|}{energy, J (meV)}& \multicolumn{2}{c|}{moment} \\ \cline{2-5}
              & U=0    & U=4eV  & U=0    & U=4eV  \\ \hline
$[010]_1$     & 267.9  & 223.7  &        &        \\ \cline{1-3}
[001]         & 33.3   & 35.5   &        &        \\ \cline{1-3}
[021]         & 3.2    & 8.9    & 4.0    & 4.4    \\ \cline{1-3}
$[010]_2$     & -12.4  & -7.6   &        &        \\ \cline{1-3}
[011]         & 19.4   & 19.8   &        &        \\ \hline
\end{tabular}
\caption{Exchange energy and magnetic moment for a Mn pair on the (110) GaAs
         surface. The subscripts "1" and "2" in the [010] direction indicate respectively the nearest
         and 2nd nearest Mn atoms along that direction.}
\end{table}

As evident from Fig.~\ref{surf_Mn}, the effect of strong
correlations, $U$, on the exchange anisotropy is rather small except
for the shortest Mn separation where correlations {\it decrease} $J$
by approximately $15\%$. As we have seen in the study of one Mn
impurity, on-site correlations have a small tendency in delocalizing
the acceptor hole around the Mn.  According to theory, the acceptor
hole is expected to mediate the exchange interaction among the
magnetic moments of Mn pairs. Except for the shortest impurity
separation, the enhanced itinerant character induced by correlations
is still of limited range in space to have a noticeable effect on
the exchange constant. Nevertheless, a close inspection of the
numerical values of $J$ (see Table~\ref{tablesurf}) shows that, with
the exception of the pair with the shortest separation, the effect
of $U$ is always to increase $J$.\footnote{At the shortest Mn
separation, the on-site Hubbard $U$ might enhance a competitive
superexchange mechanism that tends to favor antiferromagnetic
coupling. Hence $J$ decreases in this case.} This is consistent with
the fact that exchange mediating-hole becomes more delocalized as a
result of on-site correlations.

Although the results shown in  Fig.~\ref{surf_Mn} are obtained in
the absence of SOI, we have checked that SOI has little influence on
$J$. For instance for the $[010]_2$ orientation $J$ increases by 1-2
meV depending on the direction of magnetization, for both $U=0$ and
$U=4$ eV.

\subsubsection{Magnetic anisotropy}

\begin{figure}[htbp]
{\resizebox{3.2in}{2.2in}{\includegraphics{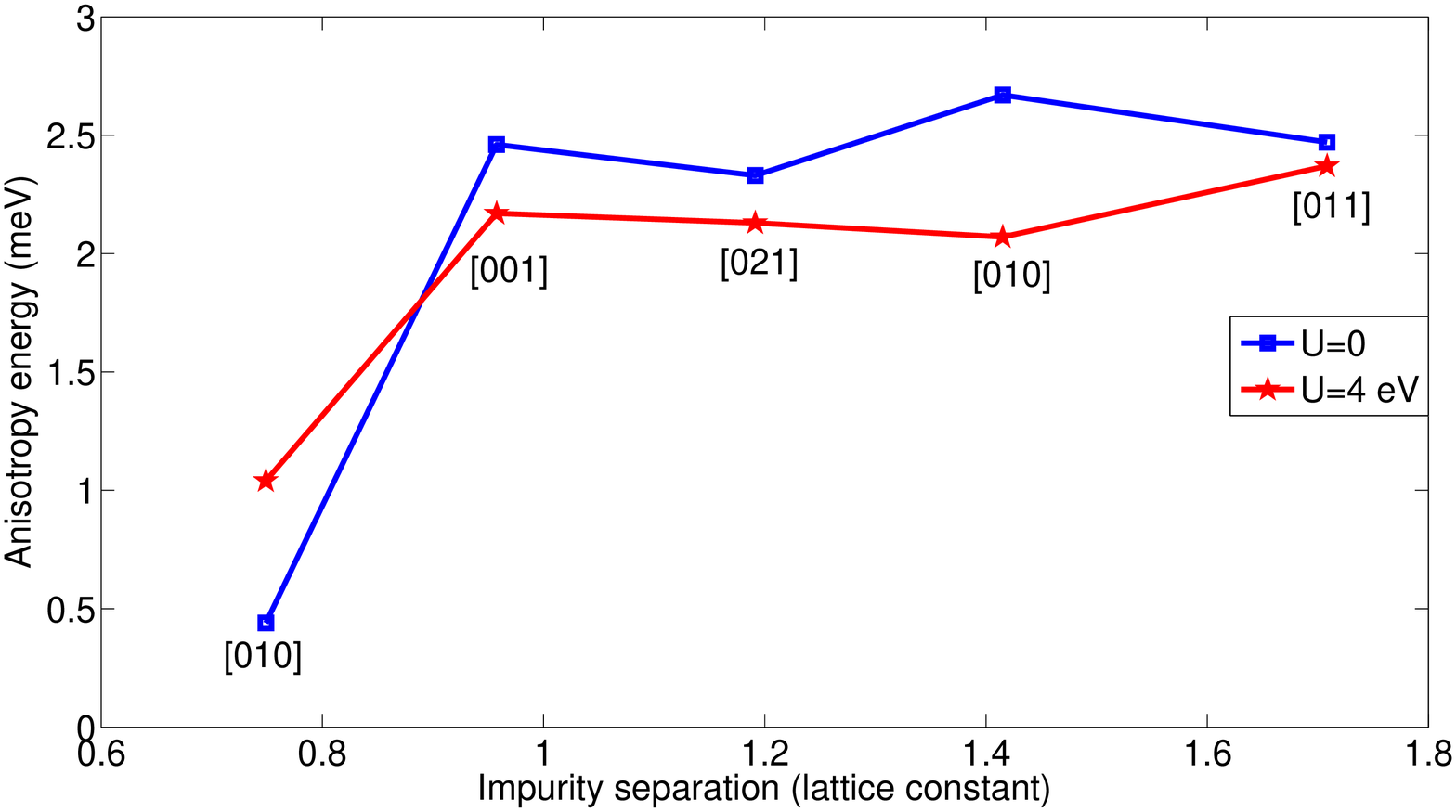}}}
\caption{(Color online) Surface anisotropy for a pair of Mn impurity
on (110) surface of GaAs.} \label{surf_aniso}
\end{figure}

The MAE barrier for a pair of Mn atom on GaAs (110) surface varies
from 0.44 meV to 2.67 meV depending on the orientation of the pair
relative to the crystallographic axes, as shown in
Fig.~\ref{surf_aniso}. For all pairs except for the pairs along the
$[010]$ orientation, the easy axis is found to be along $<001>$ and
the hard axis is along $<010>$ direction. For the  pair oriented
along $[010]$, while for the nn the situation is inverted (i.e., the
easy axis is along $<010>$ and the hard direction is along $<001>$),
for the next neighbor pair the easy axis is still along $<001>$ but
the hard axis changes to the $<100>$ direction.

Our calculations also show an interesting correlation between the MAE
and the exchange constant $J$. Comparing Figs.~\ref{surf_Mn} and
\ref{surf_aniso} we note that the Mn pair with the shortest Mn
separation, which has the largest $J$, is also the one with the
smallest MAE. For all other Mn pairs for which $J$ fluctuates around
a considerably smaller value, the corresponding MAE fluctuates around
2 meV. It is remarkable that exactly the same behavior is found in
tight-binding calculations,\cite{canali10} where the value of the
anisotropy for a Mn pair is also in quantitative agreement with the
results presented here.

In the presence of correlations $U=4$ eV, the MAE, in general,
decreases except for the case when for the pair with the shortest Mn
separation. This behavior is again consistent with the fact that,
with the exception of the shortest pair separation, on-site
correlations slightly increase the exchange constant $J$.

\section{Conclusions}
\label{conclusions_section}

In this work we have investigated the electronic and magnetic
properties of substitutional Mn impurities on the (110) surface of
GaAs, using density functional theory implementing a full potential
LAPW+lo method.

A key point of our analysis has been the investigation of of the Mn
acceptor hole state, which appears in the GaAs energy gap. The
properties of the hole state allow us to understand the salient
features of a single Mn impurity in GaAs. As for a bulk Mn impurity,
when the Mn is on the surface, the associated hole state is strongly
anisotropic in space, but much more localized around the Mn and its
two As nn than in bulk.

On site electron correlations on the Mn, included within a GGA + U
framework, reduce the hybridization of the hole state with the Mn d
orbitals, and render the character of the wave function relatively
more $p$-like and therefore slightly more delocalized. The spin
magnetic moment of the Mn atom is approximately $4 \mu_{\rm B}$,
when $U=0$, consistent with a hole partly localized on the impurity.
The Mn d moment increases to $4.5 \mu_{\rm B}$ when $U = 4$ eV, as a
result of the more delocalized nature of the hole state. However the
spin moment of the whole unit cell remains closer to $4 \mu_{\rm
B}$.

SOI induces a small dependence of the $p$ partial DOS at $E_{\rm F}$
on the direction of the magnetization, primarily for the As atoms
closest to the Mn. With some caveats (see discussion at the end of
Sec.~\ref{mae}), this dependence accounts for a calculated magnetic
anisotropy energy (MAE) of approximately 1 meV and can be closely
related to the anisotropy of the hole state.

The presence of SOI yields a small orbital magnetic moment in the
opposite direction of the spin moment, which can again be associated
to the orbital moment of the hole state. The dependence of the As
$p$ partial DOS on magnetization directions induces a very weak
anisotropy in the orbital moment. We have seen that the Bruno
formula relating the MAE to the orbital anisotropy is qualitatively
satisfied, but underestimates the MAE.

When two Mn impurities are substituted for two Ga atoms on the (110)
surface we find that the most energetically stable configuration is
typically the one where the two Mn magnetic moments are oriented
parallel to each other. The exchange energy $J$ between two Mn
impurities on the surface strongly depends on the orientation of the
pair relative to crystallographic axes. On the other hand, the
SOI-induced dependence $J$ on the direction of the magnetization is
very small. The effect of on-site correlations on $J$ is largest for
the pair along the $[010]$ direction with nn Mn,
where the exchange is reduced by 15\%. For all the other pairs,
correlations always enhance exchange, although the effect is
typically a few percents. Finally, the MAE of a ferromagnetic Mn
pair on the GaAs surface is slightly larger than that of a single
impurity, except for the shortest pair for which the MAE is 0.2
meV/Mn. With $U$ added, the MAE decreases in general.

The results presented here are in good agreement with recent
tight-binding calculations carried out for individual and pairs of
Mn substitutional impurities on the (110) surface of
GaAs.\cite{canali09, canali10, canali11} In particular, the MAE of a
single impurity is in both cases on the order of 1 meV. Similarly
the exchange energy for a Mn pair displays the same rapid decay with
Mn separation and anisotropic behavior with pair orientation.

The orbital magnetic moment of the acceptor hole is very small, as a
result of the surface that breaks the tetragonal symmetry of bulk
GaAs, again consistent with tight-binding
calculations.\cite{canali11} This suggests that for a Mn impurity on
the surface the ``total spin'' of the magnetic center composed of
the Mn core and the associated spin-polarized acceptor hole should
be $S_{\rm Tot}= 5/2 -1/2 = 2$, resulting from antiferromagnetic
coupling between the Mn spin ($S_{\rm Mn} = 5/2$) and the hole spin
($s_{\rm h}$). If this ``spin'' is subject to an anisotropy
landscape with energy barriers of 1 meV, it would take a magnetic
field on the order of 10 T to revert its direction. Although for a
Mn on surface the acceptor wavefunction is rather insensitive to the
direction of the Mn magnetic moment, this does not seem to be the
case for impurities located in the first 10 sub-layers below the
surface. In particular, the acceptor partial DOS on (110) surface
accessible from STM experiments is predicted to depend strongly on
the direction of the magnetic moment. This would imply the
possibility of controlling the STM tunneling current by rotating the
Mn acceptor with a magnetic field. So far experiments carried out
with magnetic fields up to 7 T do not find any dependence of the
conductance for different directions of the magnetic field.
\footnote{P. M. Koenraad, private communication.} If the MAE barrier
of the impurity is indeed 1 meV or larger, such fields might simply
be not strong enough to rotate the direction of the spin.

\section{Acknowledgments}

We would like to thank A. H. MacDonald, P. M. Koenraad, M. E.
Flatt{\'e}, O. Eriksson and J. Gupta for useful discussions. This work was
supported by the Faculty of Natural Sciences at Linnaeus University,
by the Swedish Research Council under Grant Numbers: 621-2007-5019
and 621-2010-3761 and by the Nordforsk research network: 08134, {\it
Nanospintronics: theory and simulations}.

\bibliography{GaAsMn}

\end{document}